\documentclass[aps,prl,amsfonts,amssymb,twocolumn,amsmath,preprintnumbers,floatfix,showpacs]{revtex4}
\usepackage{bm}
\usepackage{amsbsy}
\usepackage[dvips]{color}
\usepackage[dvips]{graphics}
\usepackage{graphicx}
\usepackage{amsmath}
\usepackage{amsbsy}
\usepackage{amssymb}
\usepackage{dcolumn}
\usepackage{bm}
\usepackage{overpic}

\DeclareMathOperator{\sgn}{sgn}

\begin{document}

\title{Low-energy subgap states and the magnetic flux periodicity\\
in $d$-wave superconducting rings}

\author{Yu.~S.~Barash}
\affiliation{Institute of Solid State Physics, Russian Academy of Sciences,
Chernogolovka, Moscow District, 142432 Russia}

\date{\today}

\begin{abstract}
Wave functions of low-energy quasiparticle subgap states in $d$-wave
superconducting rings, threaded by Aharonov-Bohm magnetic flux, are
found analytically. The respective energies are closest to the
midgap position at small magnetic fluxes and deviate from the Fermi
surface due to the Doppler shift, produced by the supercurrent.
The Doppler-shifted zero-energy states result in a paramagnetic
response of the ring at small fluxes. The states exist only for even
angular momenta of the center of mass of Cooper pairs, in agreement
with recent numerical studies of the problem. This macroscopic
quantum effect in $d$-wave rings results in broken $h/2e$
periodicity, retaining only the $h/e$ periodic behavior of the
supercurrent with varying magnetic flux.
\end{abstract}

\pacs{74.20.Rp, 74.78.Na, 74.25.Ha}

\maketitle

The $h/2e$ periodic dependence of the current on the magnetic flux in
superconducting rings and hollow cylinders is considered frequently
as an inevitable consequence of the electronic pairing in
superconductors. This point of view is fully supported by the
Ginzburg-Landau theory, which contains only the Cooper pair charge
$2e$ and always predicts the magnetic flux period $h/2e$. Numerous
experimental results seem to be also in favor of such conclusions,
including first observations of the $h/2e$ flux quantization in
hollow cylinders \cite{Doll,Deaver}, the Little-Parks effect
\cite{Parks} and the flux quantization of Abrikosov vortices
\cite{Abrikosov,Essmann}. However, the Ginzburg-Landau approach
applies on the scale much greater than the Cooper pair size and,
strictly speaking, at temperatures near $T_c$. It is, in general, not
applicable to mesoscopic rings at low temperatures, where a
microscopic approach, which is not so plain with respect to the
problem \cite{Schrieffer64,deGennes66,Loder07.1}, has to be explored.

According to the microscopic BCS theory, normal-metal electrons with
projections $\hbar M$ and $\hbar\overline{M}$ of their orbital
angular momenta along the ring axis, form in $s$-wave superconductors
Cooper pairs with the angular momentum $\hbar q=\hbar(M+\overline{M})
$ of their center of mass \cite{Bohr62,Schrieffer64}. Respectively, a
Bogoliubov quasiparticle in the superconducting ring represents a
superposition of an electron with the angular momentum $\hbar M$ and
a hole with $-\hbar\overline{M}$, and the difference is $\hbar q$.
Since $M$ and $\overline{M}$ have to be integers and can be
represented under the conditions in question as $M=(\ell+q)/2$ and
$\overline{M}=(-\ell+q)/2$, the quantities $\ell$ and $q$ can take
simultaneously either even or odd integral values
\cite{Bohr62,Schrieffer64}. The key point is that the single
valuedness of the wave functions is protected by the quantities
$(\ell\pm q)/2$, whereas only $q/2$ enters the expression for the
supercurrent via a standard combination $q/2-\Phi/\Phi_0$ with the
magnetic flux. As a result, the factor $q-2\Phi/\Phi_0$ arises in the
microscopic BCS theory and leads to the magnetic flux quantum
$\Phi_0/2$ in superconductors, a half of the normal-state flux
quantum $\Phi_0=h/e$ \cite{Byers61,Onsager61,Brenig61}.

The condition of the single valuedness of the wave function, which
justifies the presence of the superconducting flux quantum
$\Phi_0/2$, does not necessarily result itself in  the $h/2e$
periodicity, however. An important additional condition, which has
to be satisfied for ensuring the periodicity, is the degeneracy of
the respective states. In general, the degeneracy takes place between
states of a superconducting ring pierced by magnetic fluxes,
which differ by an integral number of $\Phi_0$, i. e. by an even
number of the superconducting magnetic flux quanta $\Phi_0/2=h/2e$. A
difference by odd numbers of the quanta is, in principle, physically
distinguished \cite{Schrieffer64,deGennes66}. This circumstance
represents a significant interest since it could lead to a
$h/e$-periodic component in behaviors of superconducting rings, which
removes the specific superconducting $h/2e$ periodicity or, at least,
makes it approximate.

On the other hand, there are microscopic arguments, which substantially
restrict possible deviations from the $h/2e$ periodicity and explain
from the microscopic point of view its numerous experimental
observations in superconducting rings. The arguments give also a
general idea for further search for the conditions, when the breaking
of the periodicity is noticeable. As this follows from the BCS
theory, the breaking could become observable, if, in carrying out the
statistical averaging, one cannot always replace the sum over states
near the Fermi surface by the respective integral. For this reason
the discreteness of the angular momentum is of crucial importance for
making it possible to distinguish between the states of a ring, which
differ by odd superconducting flux quanta. For $s$-wave pairing, this
can happen in mesoscopic rings with characteristic sizes of the order
of the coherence length $\xi_0$ or less \cite{deGennes66}. Among
various possible complications of the standard BCS approach which
arise in rings of such a small size, probably, the main one is that
the discreteness of the states and the pair breaking effect of the
supercurrent can destroy the superconductivity there, at least,
within some range of the magnetic flux \cite{Czajka05}. This
complicates an experimental observation of the magnetic flux
periodicity in a superconducting state of the mesoscopic $s$-wave
rings. The fluctuations of the type of quantum phase slips, as well
as effects of the Coulomb blockade, could also play an important role
there.

The problem has been investigated recently in
Ref.\onlinecite{Loder07.1} for $d$-wave rings at low temperatures,
where a noticeable violation of the $h/2e$ periodicity, associated
with discrete current-carrying low-energy states, was identified.
One of the most striking results of Ref.\onlinecite{Loder07.1}, obtained
within numerical selfconsistent calculations, is
that in $d$-wave superconducting rings quasiparticle subgap states
with quite low energies exist only for even $q$. By contrast, a
considerable empty spectral gap appears in $d$-wave rings with odd
$q$. In the presence of low-energy states, a comparatively
small variation of the Doppler shift with the magnetic flux turns out
to be sufficient for generating an alternation of the diamagnetic and
paramagnetic states in the ground state of the $d$-wave
ring. As a result, the breaking of the $h/2e$ periodic dependence of
the supercurrent on the magnetic flux has been found to be observable
in $d$-wave rings, whose size can significantly exceed
the coherence length $\xi_0$.

In the present paper an analytical approach will be developed for
studying the effect, based on a comparatively simple model of the
$d$-wave superconducting ring. Energies and wave functions of
quasiparticle current-carrying states, lying closest to the Fermi
level at small magnetic fluxes through the rings, will be obtained
by solving the respective quasiclassical equations. The
states turn out to deviate from the zero energy (the midgap position)
only due to the Doppler shift, produced by the supercurrent. They form
a paramagnetic current, which dominates the response of the ring at
low fluxes. Since the anisotropic $d$-wave pairing breaks the
conservation of the angular momentum along the ring axis,
the angular dependence of the respective
quasiparticle wave functions does not reduce to a simple exponential
dependence $\exp[i(\ell\pm q)\varphi/2]$, which they would possess in
isotropic $s$-wave or normal-metal rings. The
probability density of the states has its maxima at the nodes of the
order parameter.

The phase factors of the quasiparticle wave functions are also modified
in $d$-wave rings. This results in a topological reason for the
low-energy quasiparticle states to survive only for even $q$.
It will be shown, that the quantity $\ell/2$ in the
phase factor is replaced by the value $M_F$ of
the angular momentum
along the ring axis of the normal-state quasiparticles at the Fermi
energy, for a given transverse channel. Since $2M_F$ is always even,
the exponential $\exp[i(2M_F\pm q)\varphi/2]$ is compatible with the
single valuedness of the superconductor wave-function only for even
$q$ and, hence, for even $\ell$. Therefore, the Doppler-shifted
zero-energy states exist only in even $q$-sectors of the pairing.
This signifies that the violation of the $h/2e$ periodicity in the
subgap spectrum of $d$-wave superconducting rings is a
macroscopic quantum effect associated with the discreteness of
the states and the specific structure of their wave functions. The
discreteness is of the universal character and should be observable
at sufficiently low temperatures, when pairings with different $q$ in
superconducting rings can be distinguished.

{\it Subgap states}
Consider a narrow strongly type II superconducting ring with a negligible
supercurrent-induced magnetic flux, where the arm width $L$ is much less
than the penetration depth $L\ll\lambda_L$ and considerably exceeds the
coherence length. The nonlocal operator for an anisotropic order
parameter in the ring can be described quasiclassically and reduced,
in the main approximation, to a local angular dependent quantity.

The dependence of the Bogoliubov amplitudes on the polar angle in
continuous circular (cylindrical) rings can be taken in the form

\begin{equation}
\left\{
\begin{aligned}
&u(\varphi)=\tilde{u}(\varphi)
\exp\Bigl[\displaystyle\dfrac{i}{2}(\ell+q)\varphi\Bigr]\enspace , \\
&v(\varphi)=\tilde{v}(\varphi)
\exp\Bigl[\displaystyle\dfrac{i}{2}(\ell-q)\varphi\Bigr]\enspace , \\
&\Delta(\varphi)=\tilde{\Delta}(\varphi)\exp\left(\displaystyle
iq\varphi\right)\enspace .
\end{aligned} \right.
\label{BdGamplitudes}
\end{equation}

Quantities $\tilde{u}$ and $\tilde{v}$ do not depend on the polar
angle $\varphi$ in isotropic normal metals and $s$-wave
superconductors. Since in $d$-wave rings the angular dependence of
amplitudes $\tilde{u}(\varphi)$ and $\tilde{v}(\varphi)$ is induced
entirely by an anisotropy of the superconducting order parameter
$\tilde{\Delta}(\varphi)$ and disappears in the normal-metal state,
the amplitudes vary comparatively slowly with changing $\varphi$.
Taking this into account, the following equations of the Andreev
type for $\tilde{u}(\varphi)$ and $\tilde{v}(\varphi)$ can be derived:
\begin{align}
&i\dfrac{\hbar {\rm v}_{F,\varphi}}{R}
\dfrac{d\tilde{u}}{d\varphi}+
\Bigl(\varepsilon-\xi-m{\rm v}_{F,\varphi}
{\rm v}_s\Bigr)\tilde{u}
-\tilde{\Delta}(\varphi)\tilde{v}=0 \, , \nonumber\\
&i\dfrac{\hbar {\rm v}_{F,\varphi}}{R}
\dfrac{d\tilde{v}}{d \varphi}
-\Bigl(\varepsilon+\xi-m{\rm v}_{F,\varphi}
{\rm v}_s\Bigr)\tilde{v}
+\tilde{\Delta}^*(\varphi)\tilde{u}=0\, .
\label{andr1}
\end{align}
Here ${\rm v}_s$ is the supercurrent velocity, which takes the form
$2mR{\rm v}_s=\hbar\,{\rm min}_q\left(q-2\Phi/\Phi_0\right)$; $\Phi$
is the external Aharonov-Bohm magnetic flux and the associated
vector-potential is $A_{\varphi}=\Phi/2\pi R$. The normal-state
quasiparticle excitation energy $\xi$ is taken with respect to the
Fermi level. For a given transverse channel, $\xi$ is a discrete
quantity, which depends on $\ell/2$ in Eqs.\eqref{andr1}, and the
effective Fermi velocity of a circular motion is defined as
${\rm v}_{F,\varphi}= \partial \xi/\partial p_\varphi
\big|_{{\mathbf p}_F}$,\, $p_{\varphi}=\hbar\ell/2 R$.

As the radius of the ring is much greater than the atomic scale $k_F
R\gg 1$, the quasiclassical approach applies to a circular
quasiparticle motion and, in particular, admits the condition
$\left|{\rm v}_{F,\varphi}\right|\gg \left|{\rm v}_s\right|$, i. e.
$\left|\ell\right|\gg \left|q-2\Phi/\Phi_0\right|$. For the maximum
value of the superfluid velocity $\left|m{\rm v}_{F,\varphi}{\rm
v}_s\right|\sim {\rm max}\left[ |\Delta(\varphi)|\right]$, one gets
$\left|{\rm v}_s\right|\sim (k_F\xi_0)^{-1} \left|{\rm
v}_{F,\varphi}\right|\ll \left|{\rm v}_{F, \varphi}\right|$. Hence,
linear terms in ${\rm v}_s$, as well as all other terms in
Eqs.\eqref{andr1}, can be considered within the quasiclassical
description as small quantities of the first order as compared with
$\varepsilon_F\tilde{u}(\varphi)$.

In $s$-wave superconductors, the amplitudes $\tilde u$, $\tilde v$
do not depend on $\varphi$ and the condition
for nontrivial solutions of Eqs. \eqref{andr1}\, results in
the quasiparticle energies
\begin{equation}
\varepsilon=m{\rm v}_{F,\varphi}{\rm v}_s\pm\sqrt{\xi^2+|\Delta|^2}
\enspace . \label{envar}
\end{equation}

The ground-state value of $q$ is known to possess the $h/2e$ period,
changing with flux as $q={\rm int}[2\Phi/\Phi_0+ \sgn(\Phi)/2]$, up
to small finite-size corrections.
Such a dependence is assumed observable below, for example,
if the system is field cooled through $T_c$ for each field value. In
general, states with different $q$ describe different metastable phases
of the superconductor \cite{Bohr62}, which usually possess large
lifetimes, even if $L\ll\lambda_L$ and the magnetic flux is not really
trapped. If pair breaking effects of the supercurrent are small, the
energies \eqref{envar} depend linearly on $q-2\Phi/\Phi_0$ and manifest
the same $h/2e$ periodic dependence.

In the absence of the supercurrent, energies \eqref{envar} coincide
with those in homogeneous superconductors.  The Doppler shift,
described by the first term in Eq.\eqref{envar},
causes some of the states to move in the subgap region with varying
magnetic flux.  As $[q-(2\Phi)(\Phi_0)]\le1/2$ for the
ground-state behavior of $q$, the energy \eqref{envar} can cross the
Fermi surface at some flux value only in small rings, when
$4R<\hbar {\rm v}_{F,\varphi}/|\Delta|$ and the
pair breaking supercurrent strongly modifies or even
destroys the superconducting state in the ring. In $s$-wave rings
of larger size, energies of the subgap states lie comparatively far
from the midgap position.

Andreev equations \eqref{andr1} can be transformed to
Riccati equation \cite{Nagai93}, which has periodic coefficients
in a doubly connected $d$-wave sample.
Numerous solutions of Eqs. \eqref{andr1} will not be discussed
analytically here, as this is mostly the problem for numerical
studies. The selfconsistent numerical solution of the respective
Bogoliubov-de Gennes equations for the tight-binding model of the
d-wave square loops is represented in Ref.\onlinecite{Loder07.1}.
There are, however, two remarkably simple degenerate continuous
solutions of Eqs.\eqref{andr1} for an intrinsically real order
parameter $\tilde{\Delta}\left(\varphi\right)$, which changes its
sign and vanishes after averaging over
the polar angle. These solutions are of special interest and can be
described analytically for any particular angular dependence $
\tilde{\Delta}\left(\varphi\right)$. For a simple model of the
$d_{x^2-y^2}$-wave order parameter, one can write in the main
quasiclassical approximation $\tilde{\Delta}\left(\varphi\right)=
\Delta_d\cos2\varphi$
after applying $\Delta$ to \eqref{BdGamplitudes}. For spatially constant
$\Delta_d$, the ``electron-like'' and the ``hole-like'' solutions
take the following form (in a more general case one should make the
substitution in Eqs. \eqref{phsuv2}, \eqref{phsuv3}
$\Delta_d\sin(2\varphi)/2\rightarrow\int\tilde{\Delta}(\varphi)d\varphi$):
\begin{equation}
\begin{pmatrix}\tilde{u}(\varphi)\\ \tilde{v}(\varphi)\end{pmatrix}=
C_1\exp\Bigl(-i\beta\varphi\Bigr)\begin{pmatrix}
\cosh\left(\dfrac{\Delta_d R}{2\hbar{\rm v}_{F,\varphi}}
\sin2\varphi\right)\\ \\i\sinh\left(\dfrac{\Delta_d R}{2\hbar
{\rm v}_{F,\varphi}}\sin2\varphi\right)\end{pmatrix} \, ,
\label{phsuv2}
\end{equation}
\begin{equation}
\begin{pmatrix}\tilde{u}(\varphi)\\ \tilde{v}(\varphi)\end{pmatrix}=
C_2\exp\Bigl(-i\beta\varphi\Bigr)
\begin{pmatrix}\sinh\left(\dfrac{\Delta_d R}{2\hbar{\rm v}_{F,\varphi}}
\sin2\varphi\right)\\ i\cosh\left(\dfrac{\Delta_d R}{2\hbar{\rm v}_{F,\varphi}}
\sin2\varphi\right)\end{pmatrix} \,  ,
\label{phsuv3}
\end{equation}
where $\beta=R\xi/\hbar{\rm v}_{F,\varphi}$,
$\varepsilon=m{\rm v}_{F,\varphi}{\rm v}_s$
and constant amplitudes $C_{1,2}$ result from the normalization.

\begin{figure}[t]
\begin{center}
\includegraphics*[width=.89\columnwidth,clip=true]{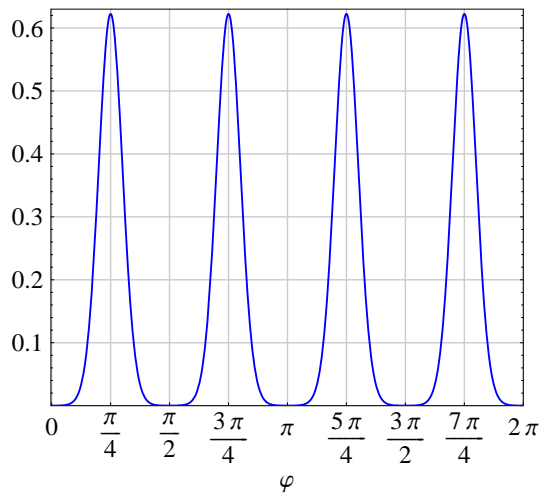}
\end{center}
\caption{The probability density as a function on the polar angle
for the Doppler-shifted zero-energy state in the d-wave ring with
$R=10\hbar{\rm v}_{F,\varphi}/|\Delta_d|$.}
\label{wavefun}
\vspace{-0.2cm}
\end{figure}

The energy $\varepsilon=\pm m\left|{\rm v}_{F,\varphi}
{\rm v}_s\right|$ deviates from
the midgap position due to the Doppler shift and its sign depends on
relative directions of the quasiparticle and the supercurrent
circulations.  If the ring size noticeably exceeds $\xi_0$, then
$|\varepsilon|\ll|\Delta_d|$. The origin of the Doppler-shifted
zero-energy states is associated with the change of sign of the
$d$-wave order parameter. Fig. \ref{wavefun} displays the probability
density of the states, which possesses the characteristic fourfold
structure and reaches its maximum exactly at the nodes $\varphi=\pm
\frac{\pi}{4},\,\pm \frac{3\pi}{4}$. As for superpositions of
\eqref{phsuv2} and \eqref{phsuv3} the fourfold symmetry reduces to
two twofold structures, the degeneracy of the states \eqref{phsuv2}
and \eqref{phsuv3}could be lifted within higher-order approximations.

The applicability of the quasiclassical approach to the solutions
\eqref{phsuv2} and \eqref{phsuv3} can be explicitly justified. The
terms containing the derivatives of the first order in Eqs.
\eqref{andr1}, can be estimated as $(\hbar{\rm v}_{F,\varphi}/R)
d\tilde{u}(\varphi)/d\varphi\sim\Delta_0\tilde{u}(\varphi)$, or
$\sim\xi\tilde{u}(\varphi)$, and identified as small quantities of
the first order. At the same time, the terms which have been
neglected within the accuracy of the Eqs.\eqref{andr1}, contain an
additional small factor $\Delta_0/\varepsilon_F$ or
$\xi/\varepsilon_F$ as compared to the first-order terms. This
concerns, in particular, the terms with the second-order derivatives,
which can be, therefore, classified as small quantities of the
second-order. A variation of the coefficient $\hbar{\rm
v}_{F,\varphi}/R$ in front of the first-order derivatives in
Eqs.\eqref{andr1}, associated with the difference between ${\rm
v}_{\varphi}$ and the respective polar component of the Fermi
velocity ${\rm v}_{F,\varphi}$, also results in small terms of the
second order, which are beyond the accuracy of the equations and have
to be neglected. This circumstance is important for further
considerations.

{\it Broken $h/2e$ periodicity}\,
Wave functions of the orbital motion have to be single-valued, if
only single-valued gauge transformations are used and spin is not
intimately involved in the problem\cite{Merzbacher62,Olariu85}. This
is the case, in particular, for the Bogoliubov amplitudes
\eqref{BdGamplitudes}, \eqref{phsuv2} and \eqref{phsuv3}. The
exponentials in Eq.\eqref{BdGamplitudes} are always single-valued,
since $\ell$ and $q$ take even or odd values only simultaneously.
Hence, the amplitudes $\tilde u(\varphi)$ and $\tilde v(\varphi)$
have to be single-valued themselves and their phases have to acquire
an integer number of $2\pi$ after going around the loop. This
signifies that the solutions (\ref{phsuv2}), (\ref{phsuv3})
really exist only if $\beta$ is an integer. For the last condition to
be satisfied, the quantization of the angular momentum along the ring
axis turns out to be of crucial importance. Indeed, close to the
Fermi surface $\left|(\ell/2)-M_F\right|\ll |M_F|$ and the quantized
normal-state excitation energy $\xi(\ell/2)$ takes the following form:
$\xi(\ell/2) ={\rm v}_{F,\varphi} (p_\varphi- p_{F,\varphi})=\frac{
\hbar{\rm v}_{F,\varphi}}{R}\left(\ell/2-M_F\right)$. Therefore,
$\beta(\ell/2)=\left(\ell/2-M_F\right)$. As $M_F$ is always an
integer, $\beta$ is an integer-valued parameter for even $\ell$ and,
hence, for even $q$. In the case of odd $\ell$ and $q$ the parameter
$\beta$ takes half-integer values, however. For this reason the
Bogoliubov amplitudes \eqref{BdGamplitudes}, \eqref{phsuv2} and
\eqref{phsuv3} are not single-valued for odd $q$, changing their sign
after going around the loop. Hence, the Doppler-shifted zero-energy
states \eqref{phsuv2}, \eqref{phsuv3} exist in the $d$-wave
superconducting rings in even $q$-sectors of the pairing, whereas for
odd $q$ the states do not arise in the gap.

As a rule, $\xi$ can be effectively excluded from a theoretical
analysis of observable superconductor properties. For example, the
current and other observables are expressed via quasiclassical Green
functions, which are $\xi$-integrated. As this follows from the
present paper, the situation can change substantially, if $\xi$ takes
discrete values due to the angular momentum quantization.

{\it Paramagnetic response}
The total current $J$ induced by the Aharonov-Bohm field
$A_{\varphi}(\rho)=\Phi/2\pi\rho$ in a cylindrical ring, is obtained
from the relation $\left(\delta{\cal E}(\Phi)\right)_{S}=-(1/c)
\int_{V} j_{\varphi }(\rho)\delta A_{\varphi}(\rho)dV=-(1/c)J\delta
\Phi$\,, where ${\cal E}$ is the energy and $S$ the entropy. Thus,
in even $q$-sectors of the pairing, a contribution to the subgap
current from the two states with $\pm\ell/2$ takes the form
$J=-\left(e {\rm v}_{F,\varphi}/2\pi R\right)\tanh\left[(\hbar {\rm
v}_{F,\varphi}/4RT) \left(q-\frac{2\Phi}{\Phi_0}\right)\right]$. In
taking the derivative of the energy over the magnetic flux, abrupt
changes associated with the $\delta q=\pm 1$ transitions have been
disregarded. Within the model, the characteristic value of the
current $e{\rm v}_{F,\varphi}/2\pi R$, carried by the low-energy
states, coincides with that for the normal-state persistent current
\cite{Ymry02}. However, its sign is determined by the quantity
$q-\frac{2\Phi}{\Phi_0}$, which is specific for the superconducting
state. As a result, it is a paramagnetic current. The current
dominates the magnetic response of the ring in vicinities of centers
of even $q$-sectors, including small fluxes at $q=0$, where
energies of the states \eqref{phsuv2}, \eqref{phsuv3} are most close
to the Fermi surface.  The paramagnetic response of the zero-energy
states is known to take place also near surfaces of $d$-wave
superconductors \cite{Barash00,Carrington01}.

{\it Conclusion}
Energies and wave functions of the low-energy quasiparticle states
have been obtained analytically in $d$-wave superconducting rings
threaded by an Aharonov-Bohm magnetic flux. The states turn out to
deviate from the zero energy due to the supercurrent-induced Doppler
shift. They form a paramagnetic response of the rings at small
fluxes. The Doppler-shifted zero-energy states are found to exist
only for pairings with even angular momenta of the center of mass of
Cooper pairs. This macroscopic quantum effect breaks $h/2e$ periodic
behavior of the supercurrent in the ring, in agreement with the
results of Ref.\onlinecite{Loder07.1}. The
analytical approach developed in the present work, demonstrates
explicitly that the quantization of the orbital angular momentum and
the condition for the single valuedness of the superconductor wave
function play the key role in a formation of the difference between
quasiparticle subgap spectra at even and odd $q$.

{\it Acknowledgements}
I am grateful to F. Lodder, A. Kampf, T. Kopp and J. Mannhart
for useful discussions. The support of RFBR grant 08-02-00842 is
acknowledged.

\end{document}